# Size Dependent Phase Transitions

Vladimir Kh. Dobruskin

Tel: (972)-506-864 642; E-mail: dobruskn@netvision.net.il

**ABSTRACT**

The contributions of heat and work in generating a new surface area are considered. Unlike the classical theory of vapor/droplet equilibrium, which associates changing surface areas with work done against the surface tension, an alternative approach assumes that the droplets grow due to a controllable condensation of vapor and the internal energy of droplets changes due to the heat of the phase transition, and not due to the mechanical work. The effect of radii on the internal energy is discussed. The theory of the vapor/droplet equilibrium is constructed on the basis of the fundamental Clapeyron equation. When droplet radii exceed about 50 Lennard-Jones' molecular diameters, the classical and new models yield similar values of thermodynamic parameters, but differ essentially in the range of the finest clusters and nanocapillaries. In contrast to the Kelvin equation, which is not applicable for adsorption in micropores with radii less than 1 nm and fails in its description of hysteresis loops in mesopores, the present approach is in reasonably good agreement with observations; the model gives rational explanations to the mechanism of a droplet growth and critical parameters of nucleation.

**Introduction**

The central position in the physical chemistry of surfaces takes a concept of a surface tension, $\gamma$, defined as follows

$$\gamma = \left(\frac{dw}{dA}\right)_{p,T} \qquad (1)$$

where $w$ is the work of reversible changing a surface area, $A$, at constant pressure, $p$, and temperature, $T$.[1,2] According to Gibbs, the interface region to a first approximation may be viewed as a two-dimensional mathematical boundary between two bulk phases both extended uniformly right up to the dividing surface.[1,2] To equalize total properties of a real system with an interface of nonzero volume with the properties of the model system, excess or surface quantities are attributed to the dividing surface. The latter is formally considered as some kind of a thermodynamic system governed by a fundamental equation, which is dependent only on the excess quantities.[1-4]

The macroscopic observations substantiate the fact that $w$ is proportional to a changeable surface area. From here the conclusion is drawn that the product $\gamma$ times $A$ ($\gamma A$-term) must be present in the fundamental equation of an interface, which postulates various modes of energy transfer between the system and the surroundings. If for a one-component bulk liquid incremental changes in the internal energy, heat, and work are connected by a basic thermodynamic equation[5]

$$dU = TdS - pdV \qquad (2)$$

where $U$, $V$ and $S$ are the internal energy, volume and entropy of the system, then for the liquid with a changeable surface area the expression takes the form[6]

$$dU = TdS - pdV + \gamma dA \qquad (3)$$

where γd*A* is the interface contribution. This term is also retained in the expression for thermodynamic potentials, which are introduced as the Legendre transformations of *U*. For example, the increment of the Gibbs free energy, *dG*, for a one-component liquid with a changeable surface area takes the form[6]

$$dG = Vdp - SdT + \gamma dA \qquad (4)$$

In the same way, γd*A*-terms are presented in the expressions for the Helmholtz free energy and the Grand potential.

Although the classical theory is limited to situations of moderately curved interfaces, it has been further modified to high curvature systems[3,4] such as nanosystems[7-10] and nucleation-growth processes.[11,12] These theories introduce supplementary terms taking into account work performed on the system by bending the interface or including allowance for the size-dependence of the surface tension. The vast literature on the theory of capillarity cannot be quoted here; our intention is only to draw attention to the role of work in generating a new area. Overall, the classical theory and its modifications consider work as a distinguishable mode of energy transfer between the system with a changeable area and the surroundings, which in principle participates in generating an interfacial area.

Contrary to this opinion, we believe that, when a surface area is changed due to the phase transitions, work is not the mode of energy transfer and the current approaches become inadequate. This conclusion relates only to the phase transitions; in other cases, unrelated to the phase transitions, such as expansion of films and soap bubbles, measurements of surface tensions, etc, work does participate in generating a new surface. In the present paper, we shall consider a contribution of heat and work and present an alternative approach to the vapor/droplet equilibrium, nucleation and growth of droplets, and adsorption in porous media.

**Mechanical and Thermal Contributions to Interface Formation**

Equilibrium thermodynamics is concerned with the idealized quasi-static processes;[13] therefore calculating a quasi-equilibrium pressure associated with uniform droplets necessitates assuming that droplets and their vapor are in equilibrium, that is, the radii and surface area are invariable. Because this assumption should be valid for any droplet radius, it means that between the state of a vapor and the state of a bulk liquid there is an infinite number of intermediate states differing with respect to particle radii. From here, the transition from nuclei of condensation to the bulk liquid may be thought of as a quasi-static locus in the thermodynamic configuration space that consists of an ordered succession of equilibrium states with growing radii. In fact, it is a controllable condensation/evaporation of vapor on/from a droplet surface. The term "droplet" implies both droplets of their own and liquid islands in pores. The effect of gravity is neglected here.

The fundamental question is how to describe a change of droplet radii. To tackle this question, we need (1)-to resort to the first law of thermodynamics

$$dU = dq + dw \qquad (5)$$

where $dq$ is a heat flow into the system, and $dw$ is work done on the system, and (2)-to examine which of these two terms, $dq$ or $dw$, is responsible for a radius change. Consider the phase transition in a container with fixed diathermal walls that is filled by the monodisperse droplets and equilibrium vapor. It is obvious that radius of droplets could be changed due to the controllable evaporation of liquid or condensation of vapor on the

5droplet surface. What are the energetic features of the process? In thermodynamics work and heat are forms of energy that are perceived by an observer *located in the surroundings* when the energy enters or leaves the system.[13, 14] It is clear that the observer perceives only the heat fluxes associated with the phase transition, as the walls are fixed and no work is done on the surroundings. Hence, the droplet/vapor interface arises due to a heat transfer; it is by no means the result of γdA-work.

The excess energy of a surface, $E_s$, relates to the surface tension as follows[2]

$$E_s = \gamma - T \frac{d\gamma}{dT} \quad (6)$$

This value is more often called the total surface energy. Here

$$S_s = -\frac{d\gamma}{dT} \quad (7)$$

is the excess surface entropy and

$$q = T\left(-\frac{d\gamma}{dT}\right) \quad (8)$$

is the quantity of latent heat absorbed in the course of the measurement of the surface tension. Notice that $E_s$ is the total surface energy, whereas γ is the free surface energy. It is clear that both $E_s$ and $U$ grow with increasing surface area, which acts as a repository for the internal energy. However, when applying the first law, the observer does not care what happens to the particular energy flow before or after it crosses the system boundary, only with the crossing process itself. *This means that the mode of energy storage in the system should not to be confused with the mode of energy transfer between the system and the surroundings.* The observer perceives a heat flow that crosses the boundary and induces a change in the internal energy.

We concentrate on the modes of energy transfer because the processes associated with work are thermodynamically nonequivalent to the processes generated by heat: a heat flux causes entropy of the system to be changed[5]

$$dq = TdS \qquad (9)$$

whereas work keeps the entropy invariable. Hence, the mechanical ($dU=dw$) and thermal ($dU=dq$) ways result in dissimilar thermodynamic states with different entropies even if $dq=dw$; these states are mapped by different points in the configuration space. In particular, consider a change of the Helmholtz free energy, $dF=dU-TdS$, at $T$=constant. The mechanical way increases the free energy ($dF>0$ if $dw=dU>0$), whereas the thermal way keeps it invariable ($dF=0$ when $dU=dq=TdS$). As a whole, the correct description of the processes generated by heat flows cannot be realized in the framework of a mechanical model, and the classical theory should be revised. Our first objective is to find correct expressions for variations of thermodynamic parameters on changing droplet radii.

**Size-Dependent Internal Energy, Enthalpy, and Entropy**

To discriminate between a bulk liquid and droplets we use the subscripts $_b$ and $_d$; the symbol $_{bd}$ will refer to the difference in properties of the bulk liquid and droplets. Consider (1)- a variation of the internal energy in the course of condensation on the surface of a bulk liquid, $\Delta U_b$, and (2)- a variation of the internal energy during the process of condensation on the surface of a droplet, $\Delta U_d$. The difference in these values ($\Delta U_b - \Delta U_d$) will be marked as $\Delta U_{db}$. Since these values will be compared at the equal temperatures, we may restrict ourselves to inspection of changes of potential energies. Because $U$ is a state function,[5] one can choose any convenient way between initial and final states for calculating $\Delta U$. In particular, imagine that condensation (see Figure 1A) consists of three consecutive stages: in the first stage, each gas molecule adsorbs on the surface (position 1) and then, in the second and third stages, it penetrates the surface layer (position 2) and, finally, into the interior of the liquid (position 3).[15-17]

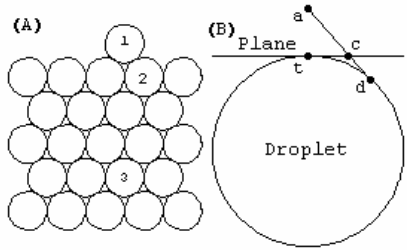

**Figure 1.** Stages of condensation. **A.** Molecule 1 absorbs on the liquid surface, molecule 2 is placed into the surface layer, and molecule 3 is in the interior of the liquid. **B.** Separations *ac* between the approaching molecule *a* and species of a flat body are less then separations *ad* between the molecule and species of the droplet. Hence, the energy of interaction with the flat body exceeds that with the droplet; it is a function of the curvature.

When the arrangement of surface layers remain invariant (e.g. for a characteristic size of more than 1 nm), the energetic effects of the second and third stages are independent of radii.[17] Nevertheless, the effect of the first stage in any case depends on *r* (Figure 1B). The first stage is autoadsorption (adsorption on the surface of own liquid). From here, the variation of the internal energy with radii is determined by the variations of the energies of autoadsorption, $\varepsilon^*$. Hence

$$\Delta U_{db} \cong -\Delta \varepsilon^*_{db} \equiv \Delta \varepsilon^*_{bd} \qquad (10)$$

As accepted in the adsorption theory, the superscript $^*$ indicates values of parameters at the well depth; besides, $\varepsilon^*$ is taken to be positive, whereas $\Delta U$ on condensation is negative. Here $\Delta\varepsilon^*_{db}=\varepsilon^*_b - \varepsilon^*_d$, where $\varepsilon^*_b$ and $\varepsilon^*_d$ are the energies of autoadsorption on the bulk liquid and on the droplet, respectively. Since for condensed phases the differences in internal energies $\Delta U_{db}$ and in enthalpies, $\Delta H_{db}$, are practically equal to one another, one has

$$\Delta H_{db} \equiv \Delta H_b - \Delta H_d \cong -\Delta\varepsilon^*_{db} \equiv \Delta\varepsilon^*_{bd} \tag{11}$$

Earlier it has been shown[15-17] that (1)-the energy of autoadsorption on the flat surface of a bulk liquid $\varepsilon_b^*$ is equal in magnitude to the total surface energy $E_s$:

$$\varepsilon^*_b = E_s \tag{12}$$

and (2)-the energy of autoadsorption on the droplet surface is equal to

$$\varepsilon^*_d = E_s \frac{\varepsilon^*_{sp}}{\varepsilon^*_{slab}} \tag{13}$$

Here $\varepsilon_{sp}^*$ and $\varepsilon^*_{slab}$ are the Lennard-Jones contributions to the energy of autoadsorption on a spherical droplet surface ($\varepsilon_{sp}^*$) and on the flat surface ($\varepsilon^*_{slab}$). These equations lead to

$$\Delta\varepsilon^*_{db} = E_s - E_s \frac{\varepsilon^*_{sp}}{\varepsilon^*_{slab}} \equiv E_s \lambda \tag{14}$$

where $\lambda$ is the geometrical factor

$$\lambda = 1 - \frac{\varepsilon^*_{sp}}{\varepsilon^*_{slab}} \tag{15}$$

The ratio $\varepsilon_{sp}^*/\varepsilon_{slab}^*$ is known from the theory of adsorption:[15]

$$\frac{\varepsilon_{sp}(R,z)}{\varepsilon^*_{slab}} = \frac{24R^3}{\sqrt{10}} \left\{ \frac{1}{\left[(R+z)^2 - R^2\right]^3} - \frac{15(R+z)^6 + 63(R+z)^4 R^2 + 45(R+z)^2 R^4 + 5R^6}{15\left[(R+z)^2 - R^2\right]^9} \right\}$$
(16)

where z is a reduced distance between a molecule and a sphere of reduced radius $R$. Note that all sizes are expressed here in the reduced forms with the Lennard-Jones diameter, $\sigma$, as a scale parameter (for example, if $r$ is the absolute radius of a droplet, then $R=r/\sigma$). For the given $R$, the extremum, $\varepsilon^*_{sp}(R, z^*)$, is found by the numerical method with respect to z; it usually occurs at $z^* \approx 0.858$. Equations 10-16 provide calculating variations of internal energy and enthalpy; a change of entropy, $\Delta S_{bd}$, with radii is given as follows:

$$\Delta S_{bd} = \frac{\Delta H_{bd}}{T} \quad (17)$$

According to the classical model, a contribution of surface area to these values is presented by

$$\Delta U = E_s \times \frac{3V_m}{r} \quad (18)$$

$$\Delta S = S_s \times \frac{3V_m}{r} \quad (19)$$

where $3V_m/r$ is the molar surface area of monodisperse droplets. Pay attention that $E_s$ in eq 18 is expressed in conventional units of $J$ per m$^2$, whereas in eqs 12-14 $E_s$ is given in $J$ per mole. The models are compared in Figure 2. Parameters of models are represented in Table 1.

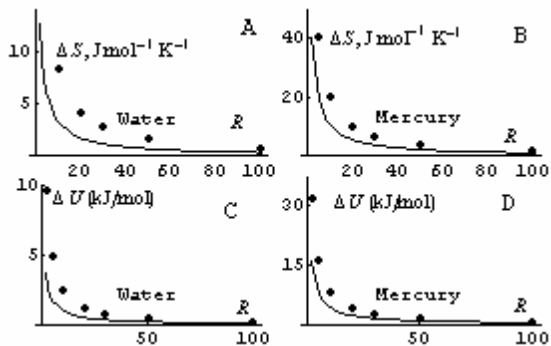

**Figure 2.** Effect of radii $R$ on entropies $\Delta S$ and internal energies $\Delta U$ of droplets. Points and a solid line refer to the classical and new models, respectively.

Although the theories are conceptually different, it is seen that for large radii (small surface areas) the models are difficult to distinguish. The difference between models increases with a reduction of $R$ and becomes visible for nanoparticles with $R<50$.

Table I. Parameters of interactions

| Substances | $T$ | $\gamma$ | $d\gamma/dT$ | $E_s$ | $\sigma$ |
|---|---|---|---|---|---|
| | (K) | (mJ m$^{-2}$) | (mJ K$^{-1}$ m$^{-2}$) | (J mol$^{-1}$) | (nm) |
| Water | 298 | 72.14[a] | -0.16[a] | 7450[a] | 0.2649[c] |
| Mercury | 393 | 448[b] | -0.162[b] | 31900[b] | 0.2898[c] |

[a] ref. 2; [b] Moelwyn-Hughes, E. A. *Physical Chemistry*; Pergamon Press: Oxford, UK, 1961.

[c] Reid R. C.; Sherwood, T. K. *The properties of Gases and Liquids*; McGraw-Hill: New York, 1958; p. 270.

**Fundamental Equation of Phase Transitions on Iinterfaces**

The fundamental Clausius-Clapeyron equation[5] takes the following forms in the cases of gas/liquid equilibriums established (1)-on the droplets:

$$\ln p_d = \Delta H_d / RT + i_d \tag{20}$$

and (2)-on the bulk liquid:

$$\ln p_b = \Delta H_b / RT + i_b \tag{21}$$

where $p_d$ and $p_b$ are the equilibrium pressures over the droplet and bulk liquid, respectively, and $i_d$ and $i_b$ are the constants of integration. Equation (20) is valid for any radii; in particular, when $r \to \infty$, $\Delta H_d$ and $p_d$ approach $\Delta H_b$ and $p_b$, respectively. Hence, $i_b = i_d$ and, subtracting eq 21 from eq 20 with eq 11 in mind, one obtains the expression

$$RT \ln \frac{p_d}{p_s} = \Delta \varepsilon_{db}^* \qquad (22)$$

Here, we take into account that $p_b$ is just the saturation pressure, $p_b \equiv p_s$. After substituting eq 14 into eq 22, the latter transforms to

$$RT \ln \frac{p_d}{p_s} = E_s \lambda \qquad (23)$$

Consider the application of the developed theory to nanodroplets. The simplest equilibrium liquid droplet consists of the central molecule and one surrounding molecular layer; the number of layers increases during the condensation. Hence, equilibrium radii can take values of 1.5, 2.5, 3.5, and so on.

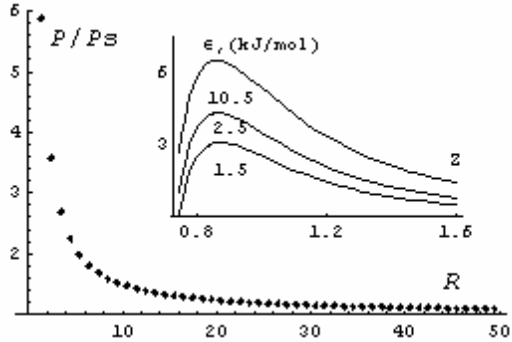

**Figure 3.** Effect of radii $R$ on the equilibrium pressures $p/p_s$ of water droplets. **Insert.** Energy of interactions (eq 14) of vapor with water droplets of $R$ 1.5, 2.5, and 10.5; $z$ is the separation between the vapor molecule and the droplet.

The energies of interactions between vapor and droplets with $R$ 1.5, 2.5, and 10.5 as a function of separation $z$ are shown in Insert (Figure 3); the maxima on the curves at $z \approx 0.858$ determine the autoadsorption energies. For a bulk water, $\varepsilon_b^* \equiv E_s = 7.45$ kJ/mol;[2]

for a droplet with $R=10.5$, $\varepsilon^*_d$ is equal to 6.5016 kJ/mol. Equilibrium oversaturations for clusters with $R=1.5$ and 2.5 are close to 6 and 4, respectively (Figure 3). A comparison of eq 22 with the Kelvin equation[15-17] showed that predicted values of $p/p_s$ are difficult to distinguish for droplets with $R>50$, but, just as for thermodynamic functions, distinctions are seen for nanodroplets with $R<20$-50.

**Equilibrium Pressures in Capillaries**

According to the classical theory, the equilibrium pressure $p$ over a concave meniscus in capillaries is given by the Kelvin equation[2, 5]

$$RT \ln \frac{p}{p_s} = -\gamma V_m \left( \frac{1}{r_1} + \frac{1}{r_2} \right) \qquad (24)$$

where $r_1$ and $r_2$ are the radii of curvature and $V_m$ is the liquid molar volume. Consider a condensation in an idealized case of a single, infinitely long, cylindrical pore. The characteristic feature of condensation in capillaries is a hysteresis loop. It is believed that the loop takes its origin from different forms of menisci in filled and empty capillaries (Insert. Figure 4).

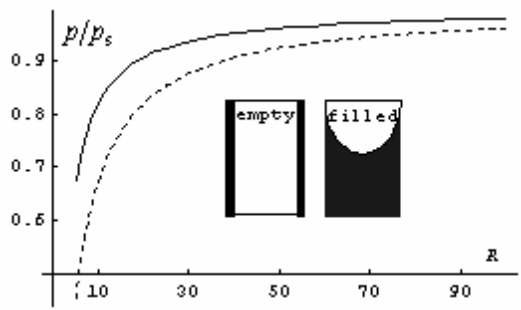

**Figure 4.** The Kelvin equation. Effect of cylinder radii $R$ on equilibrium pressures $p/p_s$ of water vapor for adsorption (solid line) and desorption (dashed line) at 293 K. **Insert**. Condensation occurs in the empty capillary when its wall is covered by the pre-adsorbed film (bold line) which takes a form of the cylinder. Desorption occurs from the filled

cylinder where the liquid surface has a form of a hemisphere.

It is seen that condensation occurs with a cylindrical meniscus, whereas evaporation (desorption) takes place from a hemispherical ones.[2, 18, 19] For a hemisphere $r=r_1=r_2$, where $r$ is the radius of capillary, while for a cylinder, $r=r_1$ and $r_2=\infty$. Substituting these values in eq 24 one obtains for the equilibrium pressures of desorption, $p_{des}$, and adsorption, $p_{ad}$:

$$RT \ln \frac{p_{des}}{p_s} = -\frac{2\gamma V_m}{r} \tag{25}$$

and

$$RT \ln \frac{p_{ad}}{p_s} = -\frac{\gamma V_m}{r} \tag{26}$$

It is seen from these equations that $p_{des}$ and $p_{ad}$ do not coincide, except the trivial case at $r=\infty$. Hence, the Kelvin equation predicts a divergence of hysteresis branches (Figure 4). But none of the porous systems conform to this model: thousands of experiments with a variety of adsorbents and adsorbates evidence that the capillary branches converge. Although many artificial assumptions[2,18, 19] were introduced, the quantitative explanations of peculiarities of hysteresis loops has yet not been given.[20] It is generally accepted to use only the desorption branch for calculating mesopore radii by eq 25 and just to ignore eq 26. Summarizing observations, Everett with co-workers[21] and Adamson[2], the authors of the fundamental studies on surface phenomena, draw attention to the fact that the situation of verification of the Kelvin equation is still conflicting!

Now consider the new model. The energy of adsorption on a cylindrical surface, $\varepsilon_{cyl}^*$:

$$\frac{\varepsilon^*_{cyl}}{\varepsilon^*_{slab}} = \frac{27}{2\sqrt{10}} \pi \left\{ \frac{21}{288R^9} \, _2F_1\left[\frac{9}{2},\frac{11}{2};1;\frac{(R-0.858)^2}{R^2}\right] - \frac{1}{3R^3} \, _2F_1\left[\frac{3}{2},\frac{5}{2};1;\frac{(R-0.858)^2}{R^2}\right] \right\} \quad (27)$$

is known from the theory[22], where $_2F_1[\alpha, \beta; \gamma; z]$ is the hypergeometric function. Substituting $\varepsilon_{cyl}^*$ and $\varepsilon_{sp}^*$ in eq 23, one has for the adsorption and desorption branches

$$RT \ln \frac{p_{ad}}{p_s} = E_s\left(1 - \frac{\varepsilon^*_{cyl}}{\varepsilon^*_{slab}}\right) \quad (28)$$

and

$$RT \ln \frac{p_{des}}{p_s} = -E_s\left(1 - \frac{\varepsilon^*_{sp}}{\varepsilon^*_{slab}}\right) \quad (29)$$

The minus sign in eq 29 arises for symmetry reasons: eq 29 relates to a concave hemisphere, whereas eq 15 refers to a convex sphere. The effect of radii on the geometrical factors for both menisci is shown in Insert (Figure 5). One may see that the geometric factors coincide at $R=2.5$; correspondingly, the adsorption and desorption branches converge at this point (Figure 5). Relative pressures of nitrogen, benzene, and water in capillaries with $R=2.5$ are equal to 0.30, 0.16, and 0.35, respectively,[15-17] in good agreement with experimental values of the beginning hysteresis loops.[19, 23, 24] Earlier, it has been shown[22] that the radii calculated by eq 28 are in accord with independent results. Hence, the new model is supported by experiments and eliminates the flaws of the Kelvin equation.

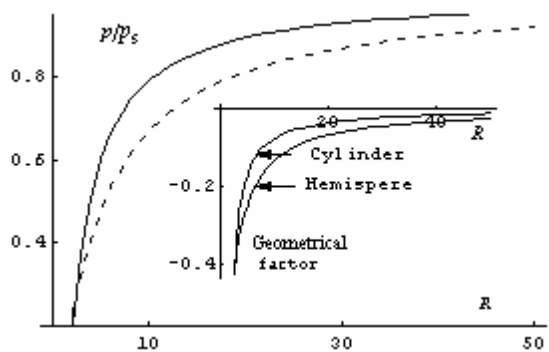

**Figure 5.** The new model. Effect of radii $R$ on equilibrium pressures $p/p_s$ for water adsorption (solid line) and desorption (dashed line) at 293 K. **Insert**. Effect of radii on the geometrical factors for hemispherical and cylindrical menisci.

**Adsorption as Evolution of Two-Dimensional Condensation**

A characteristic feature of adsorption on a homogeneous flat surface is a two-dimensional condensation (2DC) that occurs at the critical condensation pressure, $p_c$.[18, 25] The 2DC makes its appearance on the isotherm curve as a sharp jump that may be approximated by a straight line normal to the $p$-axis (Figure 6A). Adsorption on a heterogeneous sample may be presented as the sum of adsorption on the individual homogeneous areas (Figure 6A. Insert). Numerous investigations have been carried out to explore adsorption within nanopores, and virtually every analytical and numerical method used for fluids (computer simulation, DFT, lattice gases, simple thermodynamic models, etc.) has been employed.[26-32]

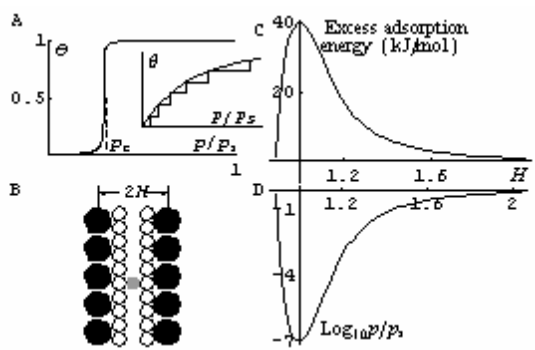

**Figure 6**. Condensation in the external field of adsorbent. **A**. Schematic isotherm of adsorption on a homogeneous surface. θ is the fractional adsorption, $p_c$ is the critical pressure of the 2DC. **Insert**. Adsorption on a heterogeneous sample may be presented as the sum of adsorption on the individual homogeneous areas. **B.** Schematic representation of a slit-like carbon micropore. The walls are formed by the carbon aromatic planes (black discs), adsorbed molecules (circles) form two adsorption monolayers; the molecule depicted as a gray disc is situated between the monolayers. $H$ is the pore half-width measured from the carbon nuclei. **C**. Effect of slit half-width $H$ on the excess energy of benzene adsorption in respect to the adsorption energy on the reference flat surface. **D.** Effect of a slit half-width $H$ on the decimal logarithm of a micropore filling pressure of benzene, $Log_{10} p/p_s$.

Because the adsorption energy is the sum of interactions with all atoms of adsorbent, it enhances in the pore due to overlapping of the field of forces of neighboring walls.[33] An adsorption equilibrium in a porous media, just as autoadsorption on droplets, is described by eq 22, but $\Delta\varepsilon^*_{db}$ is substituted by the difference between the energy of adsorption in the pore and the energy of adsorption on the reference flat surface. Equation 22 for adsorption on porous solids has been derived in many ways: (i) from the isotherm of Fowler-Guggenheim and from the quasi-chemical approximation,[34] (ii) from general thermodynamics[35] and (iii) from statistical mechanics.[35]

For a slit-like micropore of activated carbons (Figure 6B) the energy of interaction, $\varepsilon_=$, is described as[36, 37]

$$\varepsilon_=(z) = \varepsilon(z) + \varepsilon(2H - z) \qquad (30)$$

where $2H$ is the pore width, $z$ is a distance from the wall, and $\varepsilon(z)$ is the energy of interactions of adsorbate with the carbon plane:[37]

$$\frac{\varepsilon(z)}{\varepsilon^*_{slab}} = \frac{10}{3}\left\{[\tfrac{1}{5}\left(\frac{\sigma}{z}\right)^{10} - \tfrac{1}{2}\left(\frac{\sigma}{z}\right)^{4}]\right\} \qquad (31)$$

The effect of slit widths on the excess energy of adsorption is shown in Figure 6C. For example, the energy of benzene adsorption on the graphite plane is equal to $\varepsilon_{slab}=40$ kJ/mol, but in the finest carbon slit-like micropores, which can accommodate only one molecule along its width, the energy is doubled (Figure 6C).[37] As a result, the equilibrium pressure of condensation calculated by eq 22 drops to $10^{-7}p_s$ (Figure 6D).[38] The micropore filling pressures calculated by eq 22 are in the quantitative agreement with those predicted by density functional theory (see Figures 10 and 11 in ref 35).

It has been shown that the 2DC on the pore walls may evolve into volume filling of the pore. This phenomenon occurs when the formation of the adsorption monolayers enhances the energy of adsorption in a residual free space.[34, 35] As an illustration, Figure 6B shows a micropore which can accommodate three adsorption layers ($H\approx1.5$). Although a molecule in the middle layer (gray disc) is at a greater distance from the solid than molecules adsorbed on the walls (circles), it finds itself in the strong field between the adsorbed monolayers, which overshadows a decrease of wall attractions.[34, 35] In this case, a point is reached in the course of 2DC where the adsorption process is energetically as favorable for an adsorbate molecule to exist between the monolayers of adsorbate, as it is to complete the monolayer coverage. At this point, the 2DC begins to evolve in volume filling of a residual space at the same critical pressure. A condensation in cylindrical capillaries proceeds through the similar mechanism: it starts as the 2DC condensation on the wall and then evolves in volume filling.[22] Therefore, $p_{ad}$ in eq 28 in fact is the critical pressure of the 2DC in the cylindrical capillary, e. g. $p_{ad}=p_c$; more details may be found in the literature.[35]

Equation 22 has been generalized to the condensation in the system of pores of random sizes. An isotherm of adsorption in such a system may be also viewed as a "stairway" (see Figure 6A. Insert), the height of each jump and its position on the axes of pressure being determined by the volume of the individual pores, their quantities, and the energy of adsorption in the pore. Such an approach to adsorption has been realized proceeding from the hypothetical normal law of distribution of pores over their dimensions; it correctly describes the experimental isotherms,[34, 35] provides a calculation of pore size distributions,[38] and allows predicting the adsorption isotherms.[35, 39]

**Nucleation and Growth of Droplets**

In this section, we are concerned with an adsorption mechanism of droplet formation; a nucleation rate is beyond the scope of our present work. Homogeneous nucleation is the first step of the phase transition. The relation between equilibrium oversaturations and radii droplets was given in Figure 3. Consider, for example, the growth of water droplets. It is obvious that spherical nanoparticles may come closer together during a random walk; the space between spheres at this moment may be viewed as a pore formed by curved liquid walls (see Insert to Figure 7A).

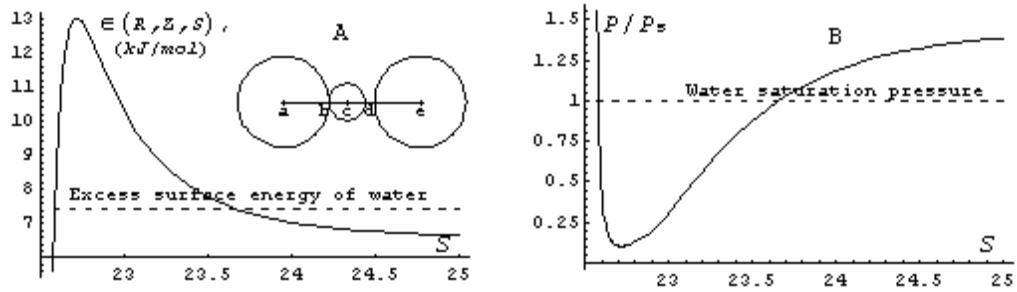

**Figure 7.** Energy of interactions, ε(R, z, s), (**A**) and equilibrium pressures $p/p_s$ (**B**) in the space between approaching water droplets of R=10.5. s is the separation between the droplets. The curves correspond to the situation where the adsorbate is located at a distance z=0.858 away from one of droplets. **Insert.** The inter-droplet space may be viewed as a pore between two droplets, a and e, the separation s=ae being the pore width. z=bc and cd=s-2R-z are distances between the molecule c and the droplets a and e.

The molecule in this "drift pore" perceives the effect of walls. For a molecule, which lies along the pore axis, the dispersion constituent of adsorption energy, ε(R, z, s), is equal to

$$\varepsilon(R, z, s) = \varepsilon_{sp}(R, z) + \varepsilon_{sp}(R, s - 2R - z) \tag{32}$$

where s is the separation between the droplets (the pore width). Here, as before, z is the distance between the molecule and the droplet and (s-2R-z) is the separation from the second sphere. The energies of interactions and equilibrium pressures $p/p_s$ in the space between approaching water droplets of R=10.5 are shown in Figure 7. When distances of adsorbate from each of droplets are close to 0.858, what amounts to s=2×(10.5+0.858) =22.716, the energy is doubled to 2×6.5016=13.032 kJ/mol (maximum in Figure 7A). Figure 7A also shows that ε(R, z, s) exceeds $E_s$ when s falls in the range 22.58 to 23.65; the equilibrium pressures in this region become smaller than $p_s$ (Figure 7B). When molecules find themselves on the periphery of droplets (to the left of *a* and to the right of *e*), the effect of overlapping vanishes, and the equilibrium pressure, as well as that for individual droplets, exceeds $p_s$ (compare with an individual droplet of $R=10.5$ in Figure 3). Hence, the molecules evaporated from the periphery of either particle will tend to condensate in the inter-droplet space, even in the unsaturated environment. This process results in the merging of the droplets and their enlargement. The model is compatible with observations that droplets are often formed on the air-solid interfaces. In these cases, the solid surfaces serve as one of the walls, while a droplet acts as the opposing wall. Since condensation in the pore leads to the droplet enlargement, the solid surface facilitates the nucleation and growth of droplets.

The maximum, doubled value of energy in the inter-droplet space, $\varepsilon^*(R, z, s)=2\varepsilon^*_{sp}$, exceeds $E_s$ only if $\varepsilon^*_{sp}>0.5E_s$. From Figure 3 follows that $\varepsilon^*_{sp}=0.4111E_s$ for $R=1.5$ and $\varepsilon^*_{sp}=0.5766E_s$ for $R=2.5$. Hence, only droplets with $R\geq2.5$ are capable of giving rise to a new phase at $p/p_s=1$. As a result, $R=2.5$ is the critical radius and such droplets manifest themselves as nucleation centers; clusters with smaller radii are incapable of growing at $p/p_s=1$. In a similar manner, the critical radii may be found at other initial pressures.

**Conclusions**

A surface area is the repository for the internal energy, which is stored as the total surface energy. In the cases where the surface area is changed due to the phase transition, the classical theory of capillarity takes the mode of energy storage for the mode of energy transfer between the system with a changeable surface area and the surroundings. A new approach develops from the assumption that the nuclei of condensation grow due to a controllable condensation of vapor on the droplet surface and a change of the thermodynamic potentials is associated with a heat flow and not with a mechanical work.

The theory of the vapor/droplet equilibrium is constructed on the basis of the fundamental Clapeyron equation. The classical and new models yield similar values of thermodynamic parameters for droplets with $R>50$, but differ essentially in the range of the finest clusters and nanocapillaries. In contrast to the Kelvin equation, which is not applicable for adsorption in micropores ($r<1$nm) and predicts unrealistic effects for hysteresis loops in mesopores ($r=1\div 50$ nm), the present approach gives a rational explanation to the peculiarities of condensation in the nanometer-size scale capillaries and describes adsorption in micropores. A general picture of adsorption, as well as predicted values of micropore filling pressures, is in a good agreement with the results of computer simulations.